\documentclass[pra,aps,showpacs,showkeys,reprint,twocolumn]{revtex4}
\usepackage{graphicx}
\usepackage{amsmath}

\begin{document}
\title{\bf Generalized Uncertainty Principle and the Conformally Coupled Scalar Field Quantum Cosmology}
\author{Pouria Pedram}
\email[Electronic address: ]{pedram@ipm.ir} \affiliation{Department
of Physics, Science and Research Branch, Islamic Azad
University, Tehran, Iran\\ School of Physics, Institute for Research
in Fundamental Sciences (IPM), Tehran 19395-5531, Iran}
\date{\today}

\begin{abstract}
We exactly solve the Wheeler-DeWitt equation for the closed
homogeneous and isotropic quantum cosmology in the presence of a
conformally coupled scalar field and in the context of the
generalized uncertainty principle. This form of generalized
uncertainty principle is motivated by the black hole physics and it
predicts a minimal length uncertainty proportional to the Planck
length. We construct wave packets in momentum minisuperspace which
closely follow classical trajectories and strongly peak on them upon
choosing appropriate initial conditions. Moreover, based on the
DeWitt criterion, we obtain wave packets that exhibit
singularity-free behavior.
\end{abstract}


 \maketitle

\section{Introduction}\label{sec1}
In recent years, many investigations have been devoted to study the
existence of a minimal length uncertainty in the context of the
Generalized (Gravitational) Uncertainty Principle (GUP)
\cite{felder}. The notion of a fundamental limit to the resolution
of structure is motivated by thought experiments as well as various
proposals of quantum gravity such as string theory, loop quantum
gravity, noncommutative geometry, and black hole physics. It
represents an approach to effectively consider quantum gravitational
aspects that cannot be treated accurately yet. This minimal length
is usually considered to be proportional to the Planck length
$\ell_{\mathrm{P}}\approx 10^{-35}m$. However, it should be noted
that as shown by Yoneya \cite{318}, the generalized uncertainty
principle is not generally valid in string theory. In particular,
the GUP introduced below is, in fact, incompatible with string
theory, and there are other approaches to quantum gravity where a
minimal length is not required.

Several testing schemes have been established to study the effects
of the quantum gravity ranging from tabletop experiments
\cite{laser4} to astronomical observations \cite{laser3,laser3-2}.
Some proposals tried to explain the puzzling observations of
ultrahigh energy cosmic rays in the framework of quantum gravity for
particle propagation \cite{laser1} and to develop high intensity
laser projects for quantum gravity phenomenology in the context of
deformed special relativity \cite{laser2}. Recently, a direct
measurement method to experimentally test the existence of the
fundamental minimal length scale was proposed using a quantum
optical ancillary system \cite{laser4}. This method is within reach
of current technology and it is based on the detection of the
possible deviations from ordinary quantum commutation relation at
the Planck scale.

The interest in deformed commutation relations started from the
seminal paper by Snyder in the relativistic framework \cite{6} and
followed by others to investigate the effects of the minimal length
on quantum mechanical and classical systems. Various problems such
as harmonic oscillator with minimal uncertainty in position
\cite{4,9} or in position and momentum \cite{7}, Dirac oscillator
\cite{11}, Coulomb potential \cite{13,pH2}, singular inverse square
potential \cite{14}, ultracold neutrons in gravitational field
\cite{24}, Lamb's shift, Landau levels, tunneling current in
scanning tunneling microscope \cite{25}, cosmological problems
\cite{Babak2,Babak3,Jalalzadeh}, and Casimir effect \cite{27} have
been treated exactly or perturbatively in the quantum domain. On the
other hand, the classical limit of the minimal length uncertainty
relation, Keplerian orbits, thermostatistics, deformations of the
classical systems in the phase space, and composite systems have
been studied at the classical level \cite{30}.

As stated above, the generalized uncertainty principle is usually
interpreted as an effective description of features of a fundamental
quantum theory of gravity. However, it is also possible to consider
the GUP as a fundamental description of nature by itself. In this
case, not only the position and momentum of a particle as variables
of quantum mechanics obey such a generalized uncertainty principle,
but also other variables should be quantized using a generalized
quantization rule. Canonical quantum gravity as an important
approach to quantize the general relativity is based on the
canonical commutation relations. Similarly, in the framework of
quantum field theory the canonical commutation relation between the
position and momentum operators is assigned to the field and its
canonical conjugated quantity. If we consider the GUP in quantum
mechanics as a fundamental characterization of nature, then the
quantum gravity and quantum field theory should be treated in the
same way. In this direction, the quantization of gravitational
fields based on the generalized uncertainty principle has been
studied in many papers \cite{many}.

The construction of wave packets in quantum cosmology as the
solutions of the Wheeler-DeWitt (WDW) equation and its relation with
classical cosmology has attracted much attention in the literature
\cite{kiefer1,tucker,sepangi,wavepacket}. These wave packets are
usually obtained by the superposition of the energy eigenfunctions
so that they follow  classical trajectories in configuration space
and peak on them whenever such classical and quantum correspondence
is feasible \cite{HALLIWELL,Matacz,HALLIWELL2}. However, since time
is absent in the theory of quantum cosmology, the initial conditions
can be formulated with respect to an intrinsic time parameter,
which, in the case of the hyperbolic WDW equation,  is taken as the
scale factor for the three-geometry \cite{DeWitt}. In particular,
the Friedmann-Robertson-Walker (FRW) quantum cosmology in the
presence of a minimally coupled scalar field is studied in
Refs.~\cite{wavepacket,pedramCQG1} and appropriate initial
conditions are presented. The quantum cosmology in a $(n +
1)$-dimensional universe with varying speed of light is investigated
in Ref.~\cite{pedramPLB3}. A class of  Stephani cosmological models
with a spherically symmetric metric and a minimally coupled scalar
field is studied in both classical and quantum domains
\cite{pedramJCAP,pedramPLB,pedramCQG2}.

In this paper, we study a closed FRW quantum cosmology with a
conformally coupled scalar field and vanishing cosmological constant
in the presence of a minimal length uncertainty. The basic results
of the conformally coupled scalar field quantum universe have been
addressed by several authors
\cite{HALLIWELL,Matacz,Schmidt,Laflamme,HALLIWELL3,Page2,Page,HALLIWELL4,Garay,Kim,Visser,kiefer2,barbosa,pedramPLB2009,barros}.
For instance, Page has studied the solutions of the WDW equation for
the FRW universe with positive, negative and zero curvatures
\cite{Page}. The WDW equation for the positive curvature transforms
into a oscillator-ghost-oscillator differential equation which is
exactly solvable. Here, both the scale factor and the scalar field
obey the modified commutation relation, i.e., $\{Q,P\}=1+\beta P^2$
where $\beta$ is the deformation parameter. We assume that the
generalized quantization principle in quantum mechanics also holds
for the quantities of quantum gravity and generalized
representations of the operators  are assigned to the observables in
the canonical approach of quantum geometrodynamics. We exactly solve
the GUP-corrected WDW equation in momentum space and construct wave
packets with classical-quantum correspondence using appropriate
choices of the initial conditions. We also obtain wave packets that
avoid singularities based on the DeWitt boundary condition.

\section{The model}\label{sec2}
Consider the Einstein-Hilbert action for the gravity and a
conformally coupled scalar field
\begin{equation}\label{action}
S=\int \mathrm{d}^4x \sqrt{-g} \left[\frac{1}{16\pi G}{\cal
R}-\frac{1}{2}\nabla_{\mu}\phi\nabla^{\mu}\phi-\frac{1}{12}{\cal
R}\phi^2 \right],
\end{equation}
where $\phi$ is the scalar field. Also, $g_{\mu\nu}$, $g$, and
${\cal R}$ denote the four-metric, its determinant, and the scalar
curvature, respectively. Units are set so that $\hbar=c=1$. The FRW
minisuperspace model with constant positive curvature and the
homogeneous scalar field are given by
\begin{eqnarray}\label{metric}
ds^2&=&-N^2(t)dt^2+a^2(t) \left[\frac{\displaystyle
dr^2}{\displaystyle 1-r^2}+r^2(d\theta^2+\sin^2\theta\,d\phi^2)
\right],\hspace{.5cm} \\  \phi&=&\phi(t).
\end{eqnarray}
Now, we define a new variable $\chi=a\ell_{\mathrm{P}}\phi/\sqrt{2}$
where $\ell_{\mathrm{P}}=\sqrt{8\pi G/3}$ and substitute
Eq.~(\ref{metric}) in the action (\ref{action}). After integrating
out the spatial degrees of freedom and discarding total time
derivatives, we find the following action \cite{Page,barbosa}:
\begin{equation}
S=\int
\mathrm{d}t\left[Na-\frac{a\dot{a}^{2}}{N}+\frac{a\dot{\chi}^{2}}{N}
-N\frac{\chi^{2}}{a}\right]. \label{action2}
\end{equation}
The corresponding Hamiltonian reads
\begin{equation}
H=N{\cal
H}=N\left[-\frac{P_{a}^{2}}{4a}+\frac{P_{\chi}^{2}}{4a}-a+\frac{\chi^{2}}
{a}\right], \label{hamiltonian}
\end{equation}
where $P_{a}=-\frac{\displaystyle2a\dot{a}}{\displaystyle N}$ and
$P_{\chi}=\frac{\displaystyle2a\dot{\chi}}{\displaystyle N}$ are the
canonical momenta conjugate to the scale factor and the scalar
field, respectively.

\subsection{The Generalized Uncertainty Principle}
In ordinary quantum mechanics the position and momentum of a
particle can be measured separately with arbitrary precision. One
way to introduce a minimal value for the measurement of position $Q$
is to modify the Heisenberg uncertainty relation to the so-called
generalized uncertainty principle as \cite{4}
\begin{eqnarray}\label{gup}
\Delta Q \Delta P \geq \frac{\hbar}{2} \left( 1 +\beta \left[(\Delta
P)^2+\langle P\rangle^2 \right]\right),
\end{eqnarray}
where $P$ is the momentum, $\beta=\beta_0/(M_{Pl} c)^2$, $\beta_0$
is of the order of unity, and $M_{Pl}$ is the Planck mass. Note that
the inequality relation (\ref{gup}) implies an absolute minimum
observable length, namely, $(\Delta Q)_{min}=\hbar\sqrt{\beta}$. In
one dimension, the above relation is given by the following deformed
commutation relation:
\begin{eqnarray}\label{gupc}
[Q,P]=i\hbar(1+\beta P^2).
\end{eqnarray}
One possible representation of this algebra is given by \cite{4}
\begin{eqnarray}\label{k1}
Q \,\psi(p)&=& i\hbar(1+\beta p^2)\partial_p\psi(p),\\ P \psi(p)&=&
p\,\psi(p),\label{k2}
\end{eqnarray}
where $Q$ and $P$ are symmetric operators on the dense domain
$S_{\infty}$ subject to the following scalar product:
\begin{eqnarray}\label{scalar}
\langle\psi|\phi\rangle=\int_{-\infty}^{+\infty}\frac{\mathrm{d}p}{1+\beta
p^2}\,\psi^{*}(p)\phi(p).
\end{eqnarray}
In this representation, the completeness relation and scalar product
read
\begin{eqnarray}\label{comp1}
\langle p'|p\rangle&=& (1+\beta p^2)\delta(p-p'),\\
1&=&\int_{-\infty}^{+\infty}\frac{\mathrm{d}p}{1+\beta p^2}\,
|p\rangle\langle p|.\label{comp2}
\end{eqnarray}
In the classical limit $\hbar\to 0$ the commutator in quantum
mechanics is replaced by the Poisson bracket for the corresponding
classical variables
\begin{eqnarray}
\frac{1}{i\hbar}[X,P]\to\{X,P\},
\end{eqnarray}
which for the deformed Heisenberg algebra (\ref{gupc}) is
\begin{eqnarray}\label{gup2}
\{Q,P\}=1+\beta P^2.
\end{eqnarray}

The nonzero uncertainty in position implies that position
eigenstates cannot be considered as physical states. This is due to
the fact that an eigenstate of an observable necessarily has
vanishing uncertainty. In fact, it is possible to construct position
eigenvectors, but these states are only formal eigenvectors and they
are not physical states \cite{4}. Notice that this feature results
from the modification of the canonical commutation relations, and
the noncommutativity of space will not, in general, imply a nonzero
minimal uncertainty. So, to obtain information on position, we
cannot use configuration space, and we need to take into account the
notion of quasiposition space.

The states that help us to recover information on positions are
maximal-localization states. These states
$|\psi^{\mathrm{ML}}_\xi\rangle$ as proper physical states have the
proprieties
$\langle\psi^{\mathrm{ML}}_\xi|Q|\psi^{\mathrm{ML}}_\xi\rangle=\xi$,
$(\Delta Q)_{\psi^{\mathrm{ML}}_\xi}=(\Delta Q)_{min}$, and obey the
minimal uncertainty condition $\Delta Q \Delta
P=|\langle[Q,P]\rangle|/2$. Thus, they satisfy the following
equation \cite{4}:
\begin{equation}
\left(Q-\langle{Q}\rangle + \frac{\langle [Q,P]\rangle}{2(\Delta P)^2}(P-\langle P\rangle)\right){\vert\psi^{\mathrm{ML}}_{\xi}\rangle}=0.
\end{equation}
In momentum space, the normalized solutions read
\begin{equation}
\psi^{\mathrm{ML}}_{\xi}(p)=\sqrt{\frac{2\sqrt{\beta}}{\pi}}(1+ \beta p^2)^{-1/2} \exp\left(-i\frac{\xi}{\hbar\sqrt{\beta}} \tan^{-1}(\sqrt{\beta}p)\right).
\end{equation}
It is now obvious that for $\beta=0$ we recover the ordinary plane
waves. Since maximal-localization states are normalizable, unlike
the canonical case, their scalar product is a function rather than a
distribution. Finally, to find quasiposition wave functions, we
project an arbitrary state $\vert\psi\rangle$ on the maximally
localized states $\vert\psi^{\mathrm{ML}}_{\xi}\rangle$, which gives
the probability amplitude for a particle being maximally localized
around the position $\xi$ with standard deviation $(\Delta
Q)_{min}$. These projections are called quasiposition wave functions
and defined by
$\psi(\xi)\equiv\langle\psi^{\mathrm{ML}}_{\xi}\vert\psi\rangle$.
Therefore, we obtain
\begin{equation}\label{ml}
\hspace{-1cm}
\psi(\xi)=\sqrt{\frac{2\sqrt{\beta}}{\pi}}\int^{+\infty}_{-\infty}\frac{dp}{(1+\beta
p^2)^{3/2}} \exp\left(i\frac{\xi}{\hbar\sqrt{\beta}}
\tan^{-1}(\sqrt{\beta}p)\right)\psi(p).
\end{equation}
For $\beta=0$, this is the well-known Fourier transformation where
the ordinary position wave functions are given by $\psi(\xi) =
\langle\xi\vert\psi\rangle$.

Before proceeding further, let us emphasize that it is difficult to
construct a meaningful position space using GUP-corrected
coordinates. Indeed, the locality problems that appear in these
approaches strongly indicate that these coordinates are not
physically relevant. However, they can be expressed in terms of the
physical coordinates that obey the standard commutation relations,
namely, $Q=(1+\beta p^2)q$ and $P=p$ where $[q,p]=i\hbar$. Here, we
study the time evolution of the GUP-corrected coordinates as well as
the physical coordinates in the classical domain. Also, in the
quantum domain, the momentum space wave function is found in terms
of the physical coordinates.

\subsection{Classical cosmology}
In the context of the GUP, the scale factor and the scalar field
satisfy the the following deformed Poisson brackets
\begin{equation}\label{poisson}
  \begin{array}{lll}
    \{a,\chi\}=0,&\hspace{1cm} & \{P_a,P_{\chi}\}=0, \\ \\
    \{a,P_a\}=1+\beta P_a^2,&\hspace{1cm}  & \{\chi,P_{\chi}\}=1+\beta P_\chi^2.
  \end{array}\displaystyle
\end{equation}
So the equations of motion for the scale factor and the scalar field
are given by
\begin{eqnarray}\label{eqm}
\left\{
  \begin{array}{ll}
    \dot{a}&=\{a,H\}=-NP_a(1+\beta P_a^2 )/(2a), \\
    \dot{P_a}&=\{P_a,H\}=2N(1+\beta P_a^2 ), \\
\dot{\chi}&=\{\chi,H\}=NP_{\chi}(1+\beta P_\chi^2 )/(2a), \\
\dot{P_{\chi}}&=\{P_{\chi},H\}=-2N(1+\beta P_\chi^2 )\chi/a.
  \end{array}
\right.
\end{eqnarray}
Using the gauge $N=a$, the time evolution of the scalar field reads
\begin{equation}
    \dot{\chi}=\{\chi,H\}=-\frac{1}{2}P_\chi(1+\beta P_\chi^2 ), \hspace{.2cm}
    \dot{P_\chi}=\{P_\chi,H\}=2\chi(1+\beta P_\chi^2 ).
\end{equation}
To proceed further, it is convenient to use the variable
$\Pi=\frac{1}{\sqrt{\beta}}\arctan\left(\sqrt{\beta}P_\chi\right)$,
which results in
\begin{equation}
    \dot{\chi}=-\frac{\tan\left(\sqrt{\beta}\Pi\right)\sec^2\left(\sqrt{\beta}\Pi\right)}{2\sqrt{\beta}}, \hspace{1cm}
    \dot{\Pi}=2\chi.
\end{equation}
So we have
\begin{equation}
    \ddot{\Pi}+\frac{\tan\left(\sqrt{\beta}\Pi\right)\sec^2\left(\sqrt{\beta}\Pi\right)}{\sqrt{\beta}}=0.
\end{equation}
If we set initial conditions so that $\chi(0)=0$ and
$P_\chi(0)=\Pi_0$, the solutions are given by
\begin{eqnarray}\label{classicsol2}
     \chi(t)&=&\frac{1}{2}\frac{\Pi_0\sqrt{1+{\eta}^2}\tan\left(\sqrt{1+{\eta}^2}\,t\right)}{\sqrt{1+\left(1+{\eta}^2\right)\tan^2\left(\sqrt{1+{\eta}^2}\,t\right)}}, \\
    P_\chi(t)&=&\frac{\Pi_0}{\sqrt{1+\left(1+{\eta}^2\right)\tan^2\left(\sqrt{1+{\eta}^2}\,t\right)}},\label{classicsol22}
\end{eqnarray}
where $\eta=\sqrt{\beta}\Pi_0$.  Also, if we set $P_a(0)=0$, the
Hamiltonian constraint $H=0$ implies $a(0)=\Pi_0/2$, and the
solutions read
\begin{eqnarray}\label{classicsol}
     a(t)&=&\frac{1}{2}\frac{\Pi_0\sqrt{1+\eta^2}\cot\left(\sqrt{1+\eta^2}\,t\right)}{\sqrt{1+\left(1+\eta^2\right)\cot^2\left(\sqrt{1+\eta^2}\,t\right)}}, \\
    P_a(t)&=&\frac{\Pi_0}{\sqrt{1+\left(1+\eta^2\right)\cot^2\left(\sqrt{1+\eta^2}\,t\right)}}.
 \end{eqnarray}
It is straightforward to check that for $\beta\rightarrow0$, the
above solutions coincide with the harmonic oscillator solutions
which represent a Lissajous ellipsis in configuration space. Note
that this solution as well as the $\beta=0$ case is singular in the
classical domain; i.e., the scale factor goes to zero as
$t\rightarrow \Pi/(2\sqrt{1+\eta^2})$.

Now, let us find the time evolution of the physical coordinates. For
our problem, these coordinates are given by
\begin{equation}
  \begin{array}{lll}
    a=(1+\beta p^2)q_a,&\hspace{1cm} & P_a=p, \\
   \chi=(1+\beta \tilde{p}^2)q_\chi,&\hspace{1cm}  & P_\chi=\tilde{p},
  \end{array}\displaystyle
\end{equation}
where $\{q_a,p\}=\{q_\chi,\tilde{p}\}=1$. Therefore, after solving
$\dot{q}_\iota=\{q_\iota,H\}$ and $\dot{p}_\iota=\{p_\iota,H\}$ for
$p_\iota\in\{p,\tilde{p}\}$, we find
\begin{eqnarray}
     q_\chi(t)&=&\frac{\Pi_0}{4}\sqrt{\left(1+{\eta}^2\right)^{-1}+\tan^2\left(\sqrt{1+{\eta}^2}\,t\right)}\nonumber\\
     &&\times\sin\left(2\sqrt{1+{\eta}^2}\,t\right), \\
    \tilde{p}(t)&=&\frac{\Pi_0}{\sqrt{1+\left(1+{\eta}^2\right)\tan^2\left(\sqrt{1+{\eta}^2}\,t\right)}},
\end{eqnarray}
and
\begin{eqnarray}
     q_a(t)&=&\frac{\Pi_0}{4}\sqrt{\left(1+{\eta}^2\right)^{-1}+\cot^2\left(\sqrt{1+{\eta}^2}\,t\right)}\nonumber\\
     &&\times\sin\left(2\sqrt{1+{\eta}^2}\,t\right), \\
    p(t)&=&\frac{\Pi_0}{\sqrt{1+\left(1+\eta^2\right)\cot^2\left(\sqrt{1+\eta^2}\,t\right)}}.
 \end{eqnarray}

\subsection{Quantum cosmology}
Let us briefly present the minisuperspace quantization of the
$\beta=0$ universe model. Following the Dirac quantization
procedure, namely, using the canonical replacement $P_a\rightarrow
-i\frac{\displaystyle\partial}{\displaystyle\partial a}$ and
$P_{\chi}\rightarrow
-i\frac{\displaystyle\partial}{\displaystyle\partial \chi}$ and
assuming a particular factor ordering, one arrives at the
Wheeler-DeWitt equation for the conformally coupled scalar field
model \cite{Page2,Page,barbosa,pedramPLB2009} ($\hbar=1$),
\begin{eqnarray}
{\cal H} \Psi(\chi,a)=
\left\{P_a^2+P_{\chi}^2+4\left(a^{2}-\chi^{2}\right) \right\}\Psi(\chi,a)\nonumber\\
=\left\{-\frac{\partial^{2}}{\partial
a^{2}}+\frac{\partial^{2}}{\partial
\chi^{2}}+4\left(a^{2}-\chi^{2}\right) \right\}\Psi(\chi,a)=0.
\label{WDW}
\end{eqnarray}
Note that this form of the WDW equation also appears in the context
of the minimally coupled scalar field \cite{pedramCQG1}, quantum
Stephani universe \cite{pedramJCAP}, and varying speed-of-light
theories \cite{pedramPLB3}. The WDW equation (\ref{WDW}) is
separable and exactly solvable in the minisuperspace variables, and
the solutions are given by
\begin{equation}
\Psi_n(\chi,a)=\psi_{n}(\chi)\psi_{n}(a), \label{4.4}
\end{equation}
where
$\displaystyle\psi_{n}(z)=\left(\frac{2}{\pi}\right)^{1/4}\left[\frac{H_{n}(
\sqrt{2}z)}{\sqrt{2^{n}n!}}\right]e^{ - z^{2}}$ and $H_{n}(z)$ is
the Hermite polynomial of degree $n$.

In the presence of the minimal length, since there are no position
eigenstates in the Heisenberg algebra representation, the Heisenberg
algebra finds no Hilbert space representation in the space of
position wave functions \cite{4}. So, we use a proper representation
of the commutation relations on wave functions in the momentum
space, i.e., Eqs.~(\ref{k1}) and (\ref{k2}). The GUP-corrected
Wheeler-DeWitt equation in momentum space now reads
\begin{eqnarray}
\Bigg\{-\left( (1+\beta p^2)\frac{\partial}{\partial p}
          \right)^2+\left( (1+\beta \tilde{p}^2)\frac{\partial}{\partial \tilde{p}}
          \right)^2\nonumber\\
          +\frac{1}{4}\left(p^{2}-\tilde{p}^{2}\right)
          \Bigg\}\Psi(p,\tilde{p})=0,\label{wdw2}
\end{eqnarray}
where we used $a=(1+\beta p^2)q_a=i(1+\beta p^2)\partial_p$,
$P_a=p$, $\chi=(1+\beta \tilde{p}^2)q_\chi=i(1+\beta
\tilde{p}^2)\partial_{\tilde{p}}$, and $P_\chi=\tilde{p}$. Notice
that $p$ and $\tilde{p}$ are physical coordinates, i.e.,
$[q_a,p]=[q_\chi,\tilde{p}]=i$. Taking
$\Psi(p,\tilde{p})=\psi(p)\psi(\tilde{p})$, we find
\begin{eqnarray}\label{diff1}
-\left[ (1+\beta p^2)\frac{\partial}{\partial p}
          \right]^2\psi(p)+\frac{1}{4}p^{2}\psi(p)=\epsilon\psi(p),
\end{eqnarray}
and a similar equation for $\psi(\tilde{p})$ where $\epsilon$ is the
separation constant. The above equation is the Schr\"odinger
equation for a simple harmonic oscillator with minimal length
uncertainty, and it can be cast into an exactly solvable
differential equation \cite{4,9,pedramPRD}. Using the new variable
$\rho=\frac{1}{\sqrt{\beta}} \tan^{-1}(\sqrt{\beta}p)$, which maps
the domain $-\infty < p < \infty$ to $-\frac{\pi}{2\sqrt{\beta}} <
\rho < \frac{\pi}{2\sqrt{\beta}}$, we find
\begin{eqnarray}
\left[\frac{\partial^2}{\partial\rho^2}-\frac{\tan^2\left(
\sqrt{\beta}{\rho}\right)}{4\beta}+\epsilon\right]\psi(\rho)=0,
\end{eqnarray}
which can be written as
\begin{eqnarray}
\left[\frac{\partial^2}{\partial\rho^2}-\frac{1}{4\beta}\frac{s^2}{
c^2}+\epsilon\right]\psi(\rho)=0,
\end{eqnarray}
where $s=\sin\left(\sqrt{\beta}{\rho}\right)$ and
$c=\cos\left(\sqrt{\beta}{\rho}\right)$. Now, we take $\psi(\rho) =
c^{\lambda}\,f(s)$, where $\lambda$ is a constant that will be
determined. So, $f(s)$ satisfies
\begin{eqnarray}
(1-s^2)f'' - (2\lambda + 1)\,s\,f' + \left[ \left\{
\frac{\epsilon}{\beta} - \lambda
         \right\}\right.\nonumber\\
         \left.
       + \left\{ \lambda(\lambda-1)-\frac{1}{4\beta^2}
         \right\}\frac{s^2}{c^2}
  \right] f
= 0,\label{fs}
\end{eqnarray}
where $-1 < s < 1$. Since the wave function should be nonsingular at
$c=0$, the coefficient of the tangent squared is required to vanish,
namely, $\lambda(\lambda-1) - \frac{1}{4\beta^2} = 0$, or
equivalently,
\begin{equation}\label{lambda}
\lambda = \frac{1}{2}\left(1+ \sqrt{1 + \frac{1}{\beta^2} }\right).
\end{equation}
Therefore, Eq.~(\ref{fs}) becomes
\begin{equation}
(1-s^2)\,f'' - (\,2\lambda + 1\,)\,s\,f' + \left(
\frac{\epsilon}{\beta} - \lambda \right) f = 0. \label{fs2}
\end{equation}
Similarly, the nonsingularity $f(s)$ at $s=\pm 1$ and requiring a
polynomial solution to Eq.~(\ref{fs}) imply
\begin{equation}
\frac{\epsilon}{\beta} - \lambda = n ( n + 2\lambda), \label{Eigen1}
\end{equation}
where $n$ is a non-negative integer \cite{SpecialFunctions}. Thus,
Eq.~(\ref{fs2}) reads
\begin{equation}
(1-s^2)f'' - (2\lambda + 1 )sf' + n ( n + 2\lambda ) f = 0.
\label{Eq:f13}
\end{equation}
The Gegenbauer polynomial is the solution of the above equation
\begin{equation}
f(s) = C_n^{\lambda}(s).
\end{equation}
Also, the energy eigenvalues are given by Eqs.~(\ref{Eigen1}) and (\ref{lambda}) as
\begin{eqnarray}\label{en}
\epsilon_n  = \left(n+\frac{1}{2}\right)\left[\sqrt{1+\beta^2}+\beta\right]+\beta n^2,
\end{eqnarray}
and the normalized energy eigenfunctions read
\begin{equation}
\psi_n(p) =
2^{\lambda}\Gamma(\lambda)
\sqrt{ \frac{ n!\,(n+\lambda)\,\sqrt{\beta} }{ 2\pi\,\Gamma(n+2\lambda) } }
\;c^{\lambda}\,C_n^{\lambda}(s),
\end{equation}
where
\begin{eqnarray}
c  = \cos\sqrt{\beta}\rho = \frac{ 1 }{ \sqrt{1+\beta p^2} }, \\ s
= \sin\sqrt{\beta}\rho = \frac{ \sqrt{\beta} p } { \sqrt{1+\beta
p^2} }.
\end{eqnarray}
Since $\psi(\tilde{p})$ also obeys Eq.~(\ref{diff1}), the solution
of the WDW equation is given by
$\Psi(p,\tilde{p})=\psi_n(p)\psi_n(\tilde{p})$.

\section{Wave packets in momentum space}\label{sec3}
\begin{figure*}
\centerline{\begin{tabular}{ccc}
\includegraphics[width=8cm]{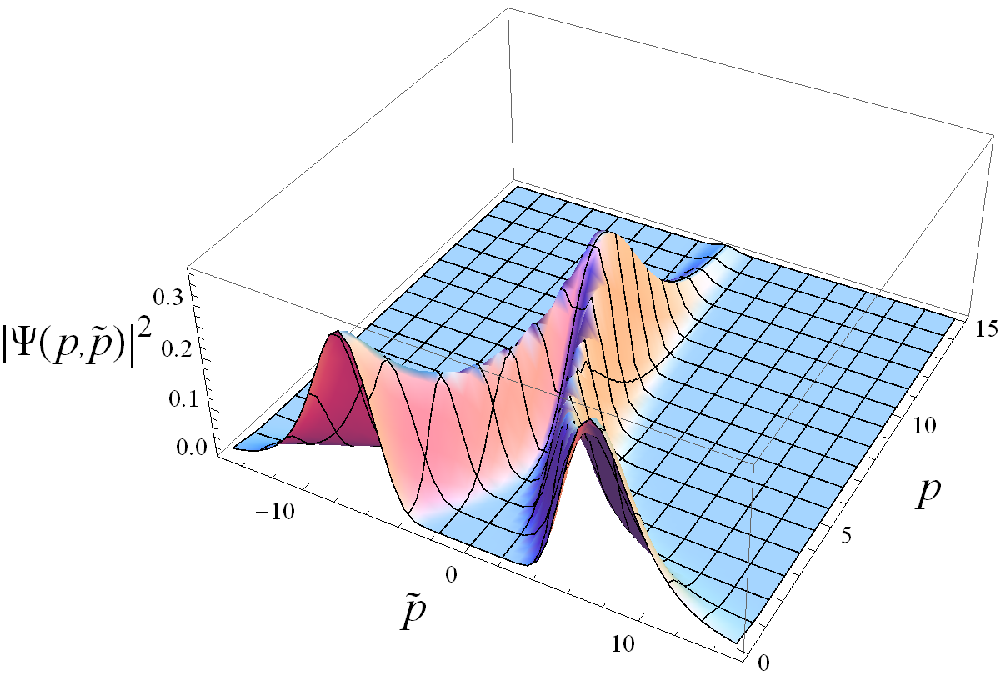}
 &\hspace{2.cm}&
\includegraphics[width=6cm]{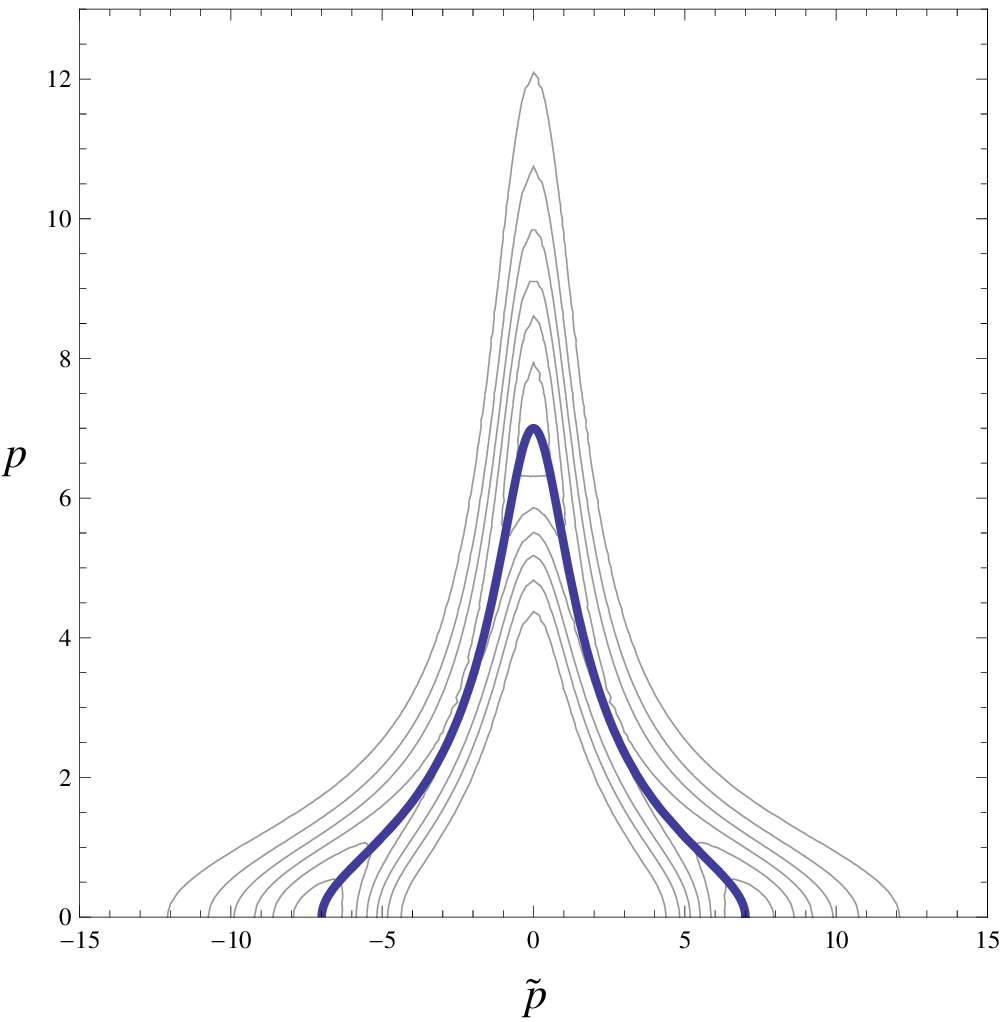}
\end{tabular}}
\caption{Left: The square of the wave packet $|\Psi(p,\tilde{p})|^2$
(\ref{wave}) for $\beta=0.1$,
$f(n)=\frac{{\,\kappa^n}}{\,{\sqrt{2^n\,n!}}}e^{-|\kappa|^2/4}$,
$\theta=0$, and $|\kappa|=4$. Right: The contour plot of the wave
packet with the classical path ($\Pi_0=7$, $\beta=0.1$) superimposed
as a thick blue line.} \label{fig1}
\end{figure*}

\begin{figure*}
\centerline{\begin{tabular}{ccc}
\includegraphics[width=8cm]{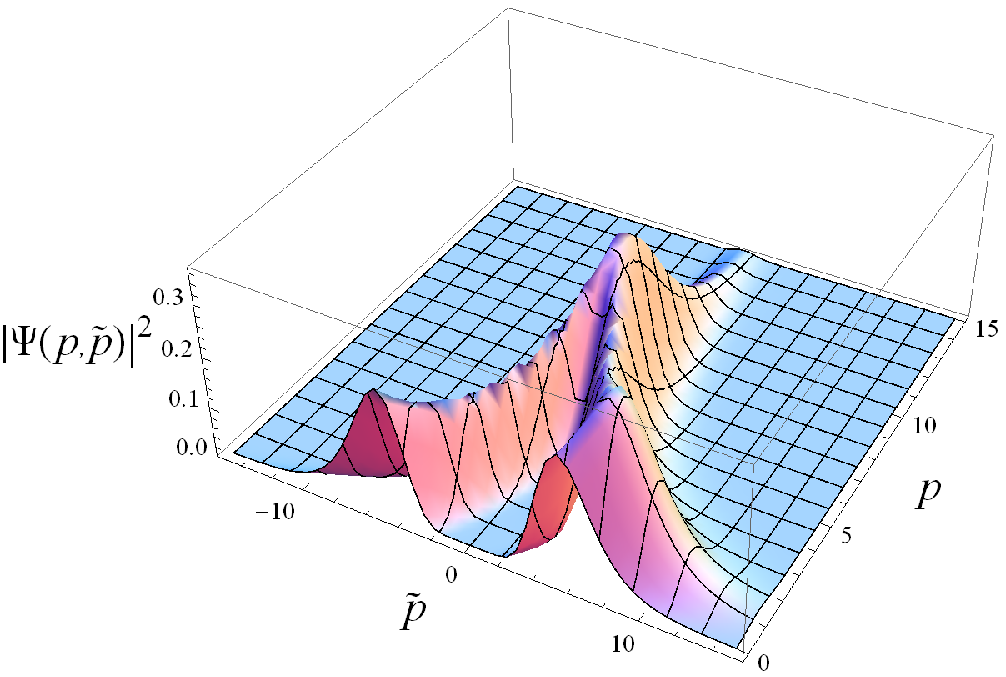}
 &\hspace{2.cm}&
\includegraphics[width=6cm]{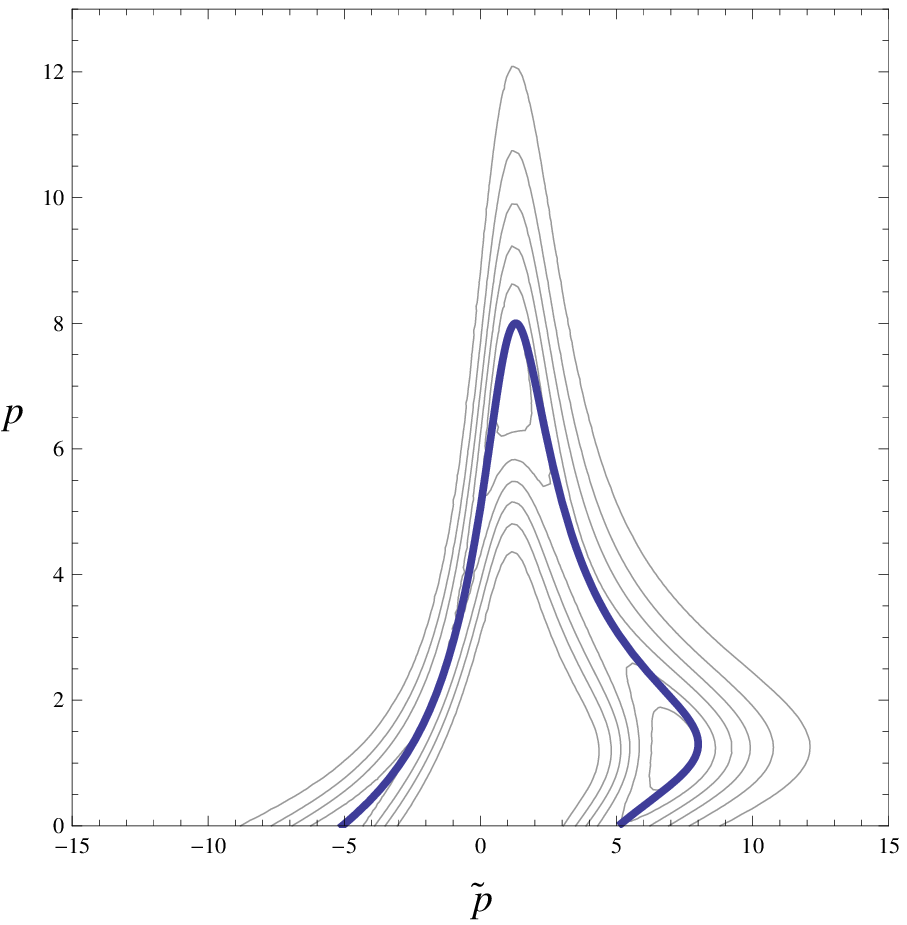}
\end{tabular}}
\caption{Left: The square of the wave packet $|\Psi(p,\tilde{p})|^2$
(\ref{wave}) for $\beta=0.1$,
$f(n)=\frac{{\,\kappa^n}}{\,{\sqrt{2^n\,n!}}}e^{-|\kappa|^2/4}$,
$\theta=\pi/8$, and $|\kappa|=4$. Right: The contour plot of the
wave packet with the classical path ($\Pi_0=8$, $\Delta=-1$, and
$\beta=0.1$) superimposed as the thick blue line.} \label{fig2}
\end{figure*}

The general solution of the WDW equation (\ref{wdw2}) is
\begin{equation}\label{psi}
\Psi(p,\tilde{p})=\sum_{n=\mbox{\footnotesize{even}}} A_n
\psi_n(p)\psi_n(\tilde{p})+i\sum_{n=\mbox{\footnotesize{odd}}} B_n
\psi_n(p)\psi_n(\tilde{p}),
\end{equation}
where the separability of the eigenfunctions into even and odd
categories is due to the parity symmetry of the WDW equation. Also,
the initial wave function and the initial derivative of the wave
function are given by
\begin{eqnarray}
\Psi(0,\tilde{p})=\sum_{n=\mbox{\footnotesize{even}}} A_n
\psi_n(0)\psi_n(\tilde{p}),\\
\frac{\partial\Psi(p,\tilde{p})}{\partial
p}\bigg|_{p=0}=i\sum_{n=\mbox{\footnotesize{odd}}} B_n
\psi_n(\tilde{p})\psi'_n(0).
\end{eqnarray}
So, $A_n$ determines the initial wave function and $B_n$  determines
its initial derivative. Notice that we are free to choose arbitrary
expansion coefficients. However, since we are interested in
constructing wave packets with classical and quantum correspondence,
these coefficients will not be independent anymore
\cite{pedramCQG1}. To address the issue of the initial conditions,
consider the WDW equation (\ref{wdw2}) near $p=0$, namely,
\begin{eqnarray}
\left\{\left( (1+\beta \tilde{p}^2)\frac{\partial}{\partial \tilde{p}}
          \right)^2-\frac{\partial^2} {\partial
p^2}- \frac{1}{4}\tilde{p}^{2}\right\}\psi(p,\tilde{p})=0.
\label{eq10nearv0}
\end{eqnarray}
The solution to this equation can be written as
$\psi(p,\tilde{p})=\xi(p)\psi(\tilde{p})$, which results in
\begin{eqnarray}
\frac{d^2\xi(p)}{d p^2}+\epsilon\,\xi(p)&=&0,
\label{eqseparated1}\\
-\left( (1+\beta \tilde{p}^2)\frac{d}{d \tilde{p}}
          \right)^2\psi(\tilde{p})+\frac{1}{4}\tilde{p}^{2}\psi(\tilde{p})&=&\epsilon\,\psi(\tilde{p}),\label{eqseparated2}
\end{eqnarray}
where $\epsilon$ is the separation constant with discrete values
(\ref{en}). Thus, the general solution to equation
(\ref{eq10nearv0}) is
\begin{eqnarray}
\psi(p,\tilde{p})&=&\sum_{n=\mbox{\footnotesize{even}}} A^*_n
\cos(\sqrt{\epsilon_n}p)
\psi_n(\tilde{p})\nonumber\\
&&+i\sum_{n=\mbox{\footnotesize{odd}}}B^*_n\sin(\sqrt{\epsilon_n}p)
\psi_n(\tilde{p}),\label{psi-separated2}
\end{eqnarray}
which is valid for small values of $p$. Now, the initial wave
function and its initial slope read
\begin{eqnarray}
\psi(\tilde{p},0)&=&\sum_{n=\mbox{\footnotesize{even}}}A^*_n\psi_n(\tilde{p}),\label{eqinitial1}\\
\psi'(\tilde{p},0)&=&i\sum_{n=\mbox{\footnotesize{odd}}}B^*_n\sqrt{\epsilon_n}\psi_n(\tilde{p}),\label{eqinitial2}
\end{eqnarray}
where the prime denotes the derivative with respect to $p$. The
prescription for choosing the coefficients is that they have the
same functional form \cite{pedramCQG1,pedramPLB3,pedramJCAP}, namely
\begin{equation}\label{eqcanonicalslope}
\left\{
\begin{array}{ll}
 A^*_n=f(n),\,\,\,\,\,\, \mbox{for $n$
  even},\\
 B^*_n=f(n),\,\,\,\,\,\, \mbox{for $n$
  odd},
   \end{array}\displaystyle
   \right.
\end{equation}
where $f(n)$ is chosen so that the initial wave function has a
proper classical description. In terms of $A_n$ and $B_n$, we obtain
\begin{equation}\label{eqcanonicalslope2}
\left\{
\begin{array}{ll}
A_n=\frac{\displaystyle
1}{\displaystyle\psi_n(0)}f(n),\,\,\,\,\,\,\mbox{for $n$
  even},\\
 B_n=\frac{\displaystyle\sqrt{\epsilon_n}}{\displaystyle\psi'_n(0)}f(n),\,\,\,\,\,\, \mbox{for $n$
  odd}.
     \end{array}\displaystyle
      \right.
\end{equation}
Now, using Eqs.~(\ref{psi}) and (\ref{eqcanonicalslope2}), the wave
packet takes the following form in momentum space
\begin{widetext}
\begin{eqnarray}\label{wave}
 \Psi(p,\tilde{p})&=&\frac{2^{\lambda}\beta^{1/4}\Gamma(\lambda)}{\sqrt{2\pi}}\left(1+\beta
p^2\right)^{-\lambda/2}\left(1+\beta
\tilde{p}^2\right)^{-\lambda/2}\Bigg\{\sum_{n=\mbox{\footnotesize{even}}}
\frac{1}{C_n^{\lambda}(0)} \sqrt{ \frac{ n!\,(n+\lambda) }{
\Gamma(n+2\lambda) } } \;f(n)\,
 C_n^{\lambda}\left(\frac{\sqrt{\beta}p}{\sqrt{1+\beta
p^2}}\right)\,C_n^{\lambda}\left(\frac{\sqrt{\beta}\tilde{p}}{\sqrt{1+\beta
\tilde{p}^2}}\right)\nonumber\\
&&
 +i\sum_{n=\mbox{\footnotesize{odd}}}
\,\,\frac{\displaystyle \sqrt{\epsilon_n}}{\displaystyle
2\lambda\sqrt{\beta} C_{n-1}^{\lambda+1}(0)}\sqrt{ \frac{
n!\,(n+\lambda) }{ \,\Gamma(n+2\lambda) } }
f(n)\,C_n^{\lambda}\left(\frac{\sqrt{\beta}p}{\sqrt{1+\beta
p^2}}\right)\,C_n^{\lambda}\left(\frac{\sqrt{\beta}\tilde{p}}{\sqrt{1+\beta
\tilde{p}^2}}\right)\Bigg\}.
\end{eqnarray}
\end{widetext}
In Fig.~\ref{fig1} we have depicted the resulting wave packet for
$\beta=0.1$ and
$f(n)=\frac{{\displaystyle\,\kappa^ne^{-|\kappa|^2/4}}}{\displaystyle\,{\sqrt{2^n\,n!}}}$,
where $\kappa=|\kappa|e^{-i\theta}$, $\theta=0$, and $|\kappa|=4$.
This choice of the coefficients is due to the fact that $f(n)$
results in the initial wave function with two well-separated peaks.
One peak corresponds to the initial values of classical momenta
($P_a,P_\chi$), and the other corresponds to their final values. As
the figure shows, the wave packet is smooth, and the crest of the
wave packet closely follows the corresponding classical trajectory.
Figure \ref{fig2} shows the wave packet corresponding to a
nonsymmetric classical trajectory, i.e., $t\rightarrow t+\Delta$ in
Eq.~(\ref{classicsol22}),
\begin{eqnarray}
    P_\chi(t)&=&\frac{\Pi_0}{\sqrt{1+\left(1+{\eta}^2\right)\tan^2\left(\sqrt{1+{\eta}^2}\,(t+\Delta)\right)}},
\end{eqnarray}
where we set $\Delta=-1$. For this case, we have $\beta=0.1$,
$f(n)=\frac{{\displaystyle\,\kappa^ne^{-|\kappa|^2/4}}}{\displaystyle\,{\sqrt{2^n\,n!}}}$,
$\theta=\pi/8$, and $|\kappa|=4$. As the figures show, for both
cases the wave packets closely follow the classical trajectories and
peak on them. Note that this behavior is due to the proper
adjustment of the expansion coefficients in Eq.~(\ref{wave}). Also,
the initial conditions $P_\chi(0)=\Pi_0$ and $P_\chi(0)<\Pi_0$
correspond to $\theta_0=0$ and $\theta_0\ne0$, respectively.

\section{The singularity problem}
Cosmological singularities are one of the puzzling phenomena in
modern physics. In fact, because of the extreme conditions at this
stage of  evolution of the Universe, the laws of physics must break
down at the singularities. In this case, the predictive power of the
theory is lost, and much efforts has been devoted to find physical
mechanisms that eliminate the offending singularities. In the
semiclassical domain, some phenomena, such as particle production,
negative vacuum stresses, and the presence of massive scalar fields,
are proposed to escape from the classical collapse predicament.
However, all these proposals violate various positive-energy
conditions of the singularity theorems. On the other hand, it is
conjectured that quantum effects can resolve this fundamental
dilemma. Now, we present three proposals for quantum singularity
avoidance and discuss the singular nature of the obtained wave
packets. We indicate that the singularity problem can be resolved
following the DeWitt criterion.

\subsection{The DeWitt boundary condition}
\begin{figure*} \centering
\centerline{\begin{tabular}{ccc}
\includegraphics[width=8cm]{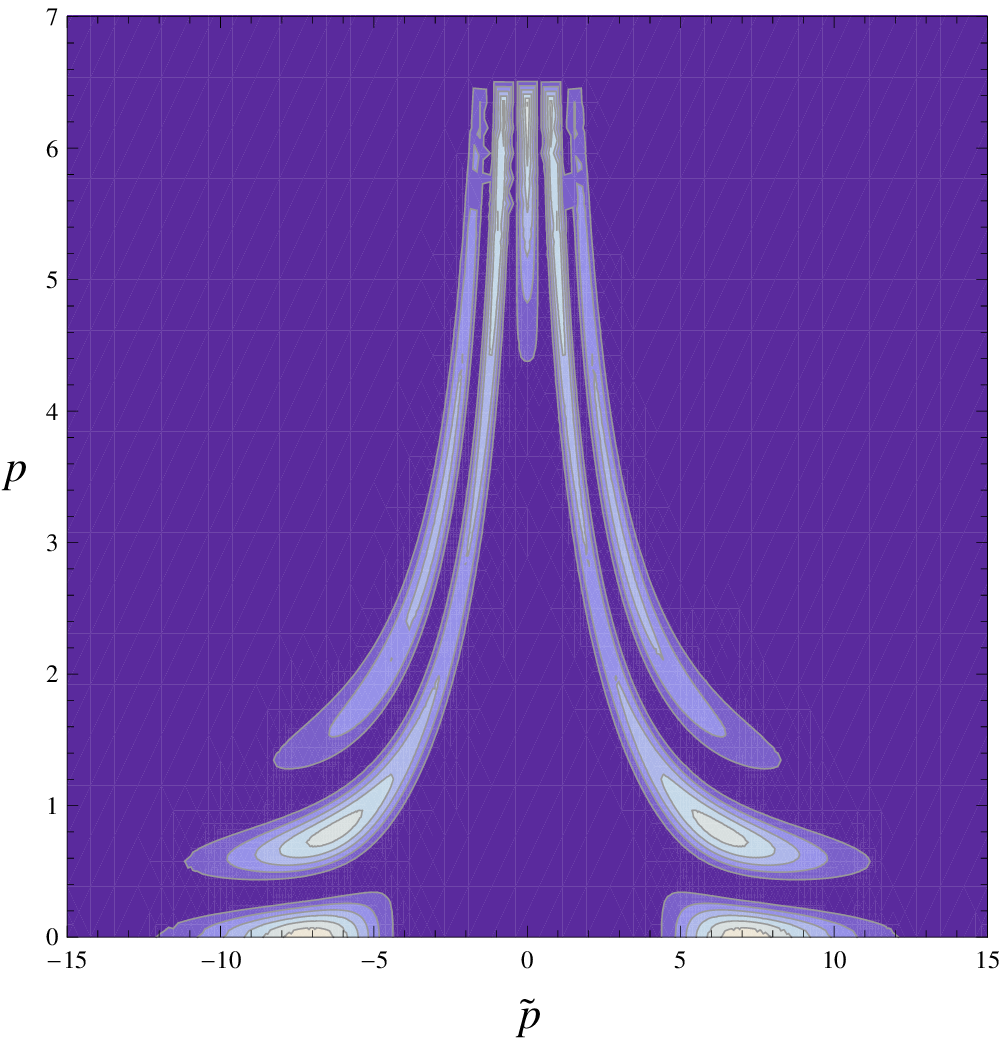}
 &\hspace{2.cm}&
\includegraphics[width=8cm]{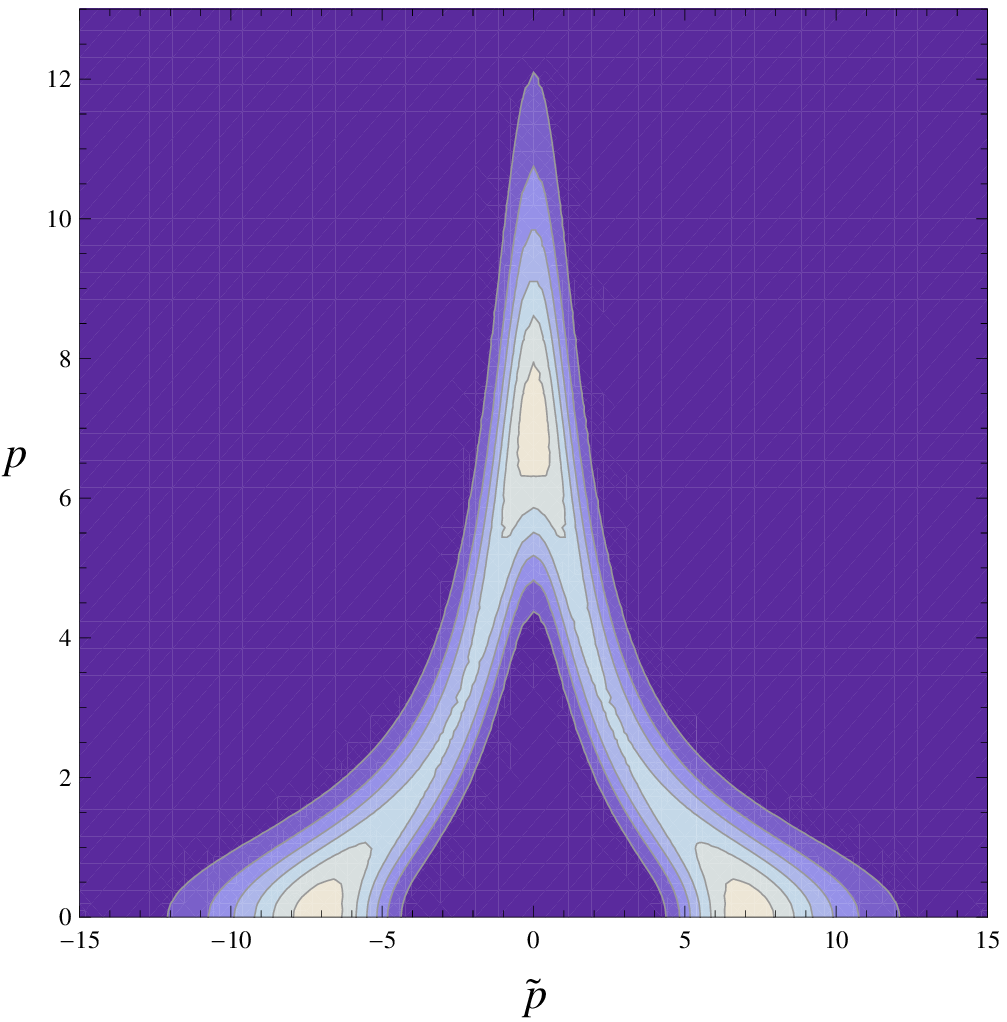}
\end{tabular}}
\caption{\label{fig3} The contour plot of the wave packet in
momentum space: Eq.~(\ref{wave2}) (left) and Eq.~(\ref{wave})
(right). We set $\beta=0.1$,
$f(n)=\frac{{\,\kappa^n}}{\,{\sqrt{2^n\,n!}}}e^{-|\kappa|^2/4}$,
$\theta=0$, and $|\kappa|=4$.}
\end{figure*}

\begin{figure}
\begin{center}
\includegraphics[width=8cm]{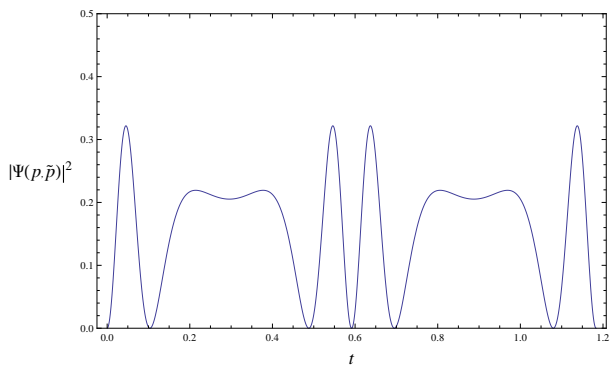}
\caption{\label{fig4} The square of the wave packet
$|\Psi(p,\tilde{p})|^2$ (\ref{wave2}) along classical trajectory for
$\beta=0.2$,
$f(n)=\frac{{\,\kappa^n}}{\,{\sqrt{2^n\,n!}}}e^{-|\kappa|^2/4}$,
$\theta=0$, and $|\kappa|=4$.}
\end{center}
\end{figure}

DeWitt suggested the following boundary condition \cite{DeWitt},
\begin{eqnarray}\label{Dcon}
\Psi\left[ ^{(3)}{\cal G}\right]=0,
\end{eqnarray}
for all three-geometries $^{(3)}{\cal G}$ related with  singular
three-geometries. So, the criterion for the quantum universe to be
singularity free is that the wave function vanishes at the classical
singularity. For our case, at fixed $\tilde{p}$, the momentum space
wave function transforms to the quasiposition wave function as
follows (\ref{ml}):
\begin{eqnarray}
\psi(\xi)\Big|_{\tilde{p}=\tilde{p}_0}&=&\sqrt{\frac{2\sqrt{\beta}}{\pi}}\int_{-\infty}^{+\infty}\frac{\mathrm{d}p}{(1+\beta
p^2)^{3/2}}\nonumber\\
&&\times
e^{\frac{i\xi}{\sqrt{\beta}}\tan^{-1}(\sqrt{\beta}p)}\Psi(\tilde{p}_0,p),
\end{eqnarray}
where $\xi=\langle a\rangle$. By taking $A_n=0$, $\Psi(p,\tilde{p})$
is an odd function of $p$ and we have $\psi(0)=0$. Consequently, the
wave function which satisfies the DeWitt boundary condition reads
\begin{widetext}
\begin{eqnarray}
 \Psi(p,\tilde{p})=
 \frac{2^{\lambda-1}\Gamma(\lambda)}{\lambda\sqrt{2\pi\sqrt{\beta}}}\left(1+\beta
p^2\right)^{-\lambda/2}\left(1+\beta
\tilde{p}^2\right)^{-\lambda/2}\sum_{n=\mbox{\footnotesize{odd}}}
\,\,\frac{\displaystyle i \sqrt{\epsilon_n}}{\displaystyle
C_{n-1}^{\lambda+1}(0)}\sqrt{ \frac{ n!\,(n+\lambda) }{
\Gamma(n+2\lambda) } }
f(n)\,C_n^{\lambda}\left(\frac{\sqrt{\beta}p}{\sqrt{1+\beta
p^2}}\right)\,C_n^{\lambda}\left(\frac{\sqrt{\beta}\tilde{p}}{\sqrt{1+\beta
\tilde{p}^2}}\right).\label{wave2}
\end{eqnarray}
\end{widetext}
This result shows that the wave packet (\ref{wave}) violates the
DeWitt boundary condition. In Fig.~\ref{fig3} we have plotted the
contour plot of the wave packet  (\ref{wave2}), which, unlike
Eq.~(\ref{wave}), it does not show the classical behavior; i.e., it
is oscillatory and vanishes at several points along a classical
trajectory (Fig.~\ref{fig4}). Note that the significance of this
boundary condition is still controversial and it has been argued
that the DeWitt boundary condition has little to do with the quantum
singularity avoidance \cite{Got2,sin1,sin2}.

\subsection{The criterion of the expectation value of observables}
Based on the proposal by Lund \cite{Lund} and Gotay and Isenberg
\cite{Got1}, a quantum state $\psi$ is singular if and only if
$\langle\psi|Qf|\psi\rangle=0$ for any quantum observable $Qf$ where
its classical counterpart $f$ vanishes at the singularity. One of
the advantages of this criterion is that it is straightforward to
check. For our case, if we write the wave function as a sum of even
and odd parts, namely $\Psi=\Psi_{\mathrm{e}}+i\Psi_{\mathrm{o}}$,
the expectation value of the scale factor observable reads
\begin{eqnarray}
\langle a\rangle&=&\langle\Psi|i\left(1+\beta
p^2\right)\frac{\partial }{\partial
p}|\Psi\rangle\bigg|_{\tilde{p}=\tilde{p}_0}\nonumber
\\ &=&i\int_{-\infty}^{+\infty}\mathrm{d}p\,\Psi^{*}(p,\tilde{p}_0)\frac{\partial
}{\partial p}\Psi(p,\tilde{p}_0)\nonumber\\
&=&\int_{-\infty}^{+\infty}\mathrm{d}p\left\{\Psi^{*}_{\mathrm{o}}(p,\tilde{p}_0)\Psi'_{\mathrm{e}}(p,\tilde{p}_0)-
\Psi^{*}_{\mathrm{e}}(p,\tilde{p}_0)\Psi'_{\mathrm{o}}(p,\tilde{p}_0)\right\},\hspace{.5cm}
\end{eqnarray}
which is identically zero for the solution (\ref{wave2}) and
vanishes at $\tilde{p}_0=0$ for the solution (\ref{wave}).
Therefore, this test for quantum collapse shows that both solutions
(\ref{wave}) and (\ref{wave2}) cannot escape the quantum mechanical
singularity. In the next subsection, we show that this result may be
due to the ``choice of time" on the classical level.

\subsection{Fast- and slow-time gauges}
It is shown that  the quantum collapse is predetermined by the
choice of time on the classical domain. In this regard, Gotay and
Demaret conjectured that slow-time quantum dynamics is always
nonsingular, while fast-time quantum dynamics is inevitably
singular, i.e., leads to the collapse \cite{Got2}. A time variable
$t$ is called fast time if the singularities occur at either
$t=-\infty$ or $t =\infty$. Otherwise, $t$ is called slow time.
Indeed, fast-time gauge dynamics is complete and can  be considered
as a regularization of a slow-time gauge dynamics which is
incomplete. This distinction is particularly useful in the quantum
mechanical domain.

For our case, classical solutions show that the scale factor runs
from $a=0$ at $t=- \Pi/(2\sqrt{1+\eta^2})$ and then collapses to
$a=0$ at $t=\Pi/(2\sqrt{1+\eta^2})$. So, the time $t$ here is slow
time. Since the above conjecture implies that the slow-time quantum
dynamics is nonsingular and as we showed before
$\langle\psi|Qf|\psi\rangle=0$ at the classical singularity, we
conclude that the quantum dynamics (\ref{wdw2}) is not quantized in
the slow-time gauge. Note that for $\beta=0$ and at the classical
level, the scale factor $a$ expands monotonically from $a = 0$ at
$\phi = - \infty$ to its maximum value and then collapses
monotonically back to $a = 0$ at $\phi = + \infty$ \cite{Page}.
Thus, if we take $t=\phi$ as a time choice, it is a fast-time gauge
which runs from $-\infty$ to $+\infty$, and its corresponding
quantum dynamics based on the effective Hamiltonian $H=-P_\phi$ will
be singular as well. For our case, although this conjecture does not
determine the time gauge of the quantum dynamics, it only indicates
that the model is quantized in a fast-time gauge.

\section{Conclusions}\label{sec5}
We have studied a closed Friedmann-Robertson-Walker quantum
cosmology model in the presence of a conformally coupled scalar
field and in the context of the generalized uncertainty principle.
In this framework, both the scale factor and the scalar field
satisfy the modified commutation relation $[Q,P]=i(1+\beta P^2)$
where $\beta$ is the GUP parameter. We exactly solved the
Wheeler-DeWitt equation in momentum space and obtained the solutions
in terms of  the Gegenbauer polynomials. In principle, since the WDW
equation is a second-order differential equation, the initial wave
function and its initial derivative, i.e., the expansion
coefficients, can be chosen freely. However, the classical and
quantum correspondence imposes a particular relation between the
expansion coefficients. Here, we proposed a particular relation
between the  even and odd expansion coefficients that determine the
initial wave function and its initial derivative, respectively. The
resulting wave packets closely followed their corresponding
classical trajectories and peaked on them in the momentum space. For
$f(n)=\frac{{\displaystyle\,\kappa^ne^{-|\kappa|^2/4}}}{\displaystyle\,{\sqrt{2^n\,n!}}}$
where $\kappa=|\kappa|e^{-i\theta}$, we showed that the symmetric
and nonsymmetric classical solutions correspond to $\theta=0$ and
$\theta\ne0$, respectively. These wave packets are also singular in
the quantum domain based on the DeWitt boundary condition. This
problem can be avoided (even in the absence of the GUP) by taking
$A_n=0$, namely, vanishing the even expansion coefficients. However,
the criterion of the expectation value of observables and the
conjecture by Gotay and Demaret showed that the singularity problem
still exists due to the fast-time gauge of the quantum dynamics.

\acknowledgments I would like to thank the referees for giving such
constructive comments which considerably improved the quality of the
paper. This research is supported by Iran National Science
Foundation (INSF) (Grant No.~92040472).

\end{document}